Non-perturbative Bacterial Identification Directly from Solid Agar Plates Using Raman Micro-spectroscopy and Machine Learning

Jeong Hee Kim[1], Jia Dong[1], Marissa Morales[1], Loza Tadesse[1,2,3,*]

[1]Department of Mechanical Engineering, MIT, Cambridge, MA, USA

[2]Ragon Institute of MGH, MIT and Harvard, Cambridge, MA, USA

[3]Jameel Clinic for AI & Healthcare, MIT, Cambridge, MA, USA

* Corresponding Author:

Loza Tadesse, PhD

Assistant Professor of Mechanical Engineering, MIT
Associate Member, Ragon Institute of MGH, MIT, and Harvard
Associate Member, J Clinic for AI and Health

**Abstract**

Raman spectroscopy is a promising tool for microbial identification, yet its implementation in microbiology and clinical workflow is still restricted due to the accompanying additional preparation required to focus on microbial signals. Here, we demonstrate Raman-based bacterial identification directly from unopened, inverted agar plates, the same conditions used during incubation. Our approach enabled identification with single gene-level sensitivity using two *Escherichia coli* variants, differing only in green fluorescent protein (GFP) expression, across diverse media and substrate material conditions, despite the interrogation path traversing 3-4 mm thick background material. We integrated traditional density functional theory (DFT)-based material computation with machine learning analysis, achieving over 97.7% classification accuracy, surpassing the performance of standard measurements from opened plates by 10.8% higher mean accuracy and 0.76% less variance. We further demonstrated Raman mapping-based colony identification via Raman peaks characteristic to GFPmut3 chromophore structure generated by DFT. Our approach is robust to changes in algorithms or substrate materials and promises real-time, non-perturbative monitoring of bacterial growth, biofilm formation, and antimicrobial resistance development.

**Introduction**

Microbial culture and identification play essential roles in various applications and sciences, ranging from clinical and pharmaceutical to food and environmental sectors[1,2]. Microorganisms can be used to produce therapeutic drugs[3,4], vaccines[5], and food products,[6] as well as to facilitate waste decomposition[7]. However, pathogenic bacteria pose risks to the environment[8] through poor water and soil hygiene and to human health[9] through contaminated food products and infections. Clinically, the rise in antimicrobial-resistant strains, often differing by only a few genes from their susceptible counterparts[10], leads to higher treatment costs, extended hospital stays, and increased mortality[11]. To prevent the outbreak of pathogenic microbes and reduce waste and delays in the production of antimicrobials or food products, rapid and accurate microbial testing modalities are vital[9]. Moreover, differentiation at the gene level is critical for selecting potent and specific antimicrobials for treatment.

Current strategies for microbial identification and susceptibility testing primarily rely on time- and labor-intensive *in vitro* cultivation followed by metabolomics monitoring by commercially available automated systems, such as VITEK 2 (bioMérieux) and Phoenix 100 (Becton Dickinson)[9]. Various agar-based media, some containing chromogenic substrate, enable the isolation and differentiation of target pathogens based on their metabolic activity, pH level, and/or enzymes[2,12]. However, they lack specificity on both species- and strain-levels, requiring additional confirmation via molecular diagnostic techniques with tedious sample preparation steps[9]. These techniques include DNA amplification-based methods, such as polymerase chain reaction (PCR) and quantitative PCR (qPCR), sequencing-based methods, such as whole genome sequencing (WGS), and matrix-assisted laser desorption/ionization time of flight (MALDI-TOF) mass spectrometry[2,13]. While enabling accurate identification, these procedures involve expensive consumables and destructive, discontinuous processes, posing challenges in reproducibility due to costly and inconvenient cross-validation from different techniques. Therefore, there is a need for rapid, reliable, and high-throughput readouts to supplement and/or replace current identification methods.

Vibrational spectroscopy has been demonstrated as an alternative approach to identify microorganisms by molecular fingerprinting[14–16]. In particular, Raman spectroscopy, owing to its label-free and non-destructive manner of measurement that is robust to water content, has been recently expanding in microbial applications[17–20]. Raman spectroscopy enables microbial identification via probing inelastic scattering unique to constituting molecules as microbes interact with monochromatic light[17,20]. Combined with machine learning, Raman spectroscopy can accurately and objectively distinguish various bacterial species with single-cell[21] and strain-level sensitivity[22], while alleviating demands for expertise in chemometrics[23]. Current implementations, however, lack integration with existing workflows for microbial culture, introducing additional preparation steps, including colony isolation, washing, and transfer to substrates for selective signal enhancement of optically transparent microbes[24–27]. To mitigate these challenges, recent studies[28–30] have demonstrated direct Raman measurement on solid agar, achieving 94.1 % accuracy in identifying 10 bacterial species[28]. While improving integration and throughput, these approaches expose microbes to the environment by opening the plates, thus endangering culture purity and operator health, in addition to technical challenges, such as limited measurement time and constant refocusing demands due to agar dehydration. To enable more integrable measurements without contamination or sample degradation risks, there is a need for a robust, high-throughput identification directly from colonies without opening culture plates.

In this study, we employed Raman micro-spectroscopy to identify bacterial colonies directly from an unopened, solid agar plate placed in an inverted orientation, the same orientation used for standard culturing and storage. Our approach enabled distinguishing *Escherichia coli* (*E. coli*) colonies grown on two commonly used agar media, lysogeny broth (LB) agar and tryptic soy agar (TSA), prepared either in a quartz-bottomed or a polystyrene (PS) petri dish. Despite differences in the background materials, we successfully recovered Raman features characteristic of bacterial colonies that are confirmed by Raman spectra acquired via the standard upright orientation. Our approach demonstrated sensitivity by objectively distinguishing colonies of *E. coli* genetically modified to express green fluorescence protein (GFP) from spectra from wild type colonies, by leveraging a synergistic combination of density functional theory (DFT) and machine learning analysis

We further demonstrate Raman mapping using our approach, which otherwise is challenging due to agar dehydration and frequent refocusing demands, thereby promising Raman-based automated colony identification and rapid, high-throughput antimicrobial susceptibility testing. The use of a 10× objective lens allows clearance of 5-6 mm between the lens and the petri dish, facilitating the repositioning of the area of interest within the sample substrate. This closed-lid, inverted measurement setup mitigates contamination concerns and allows for additional measurements or downstream incubation, opening up the potential for real-time monitoring of bacterial growth and antimicrobial resistance.

**Results**

*Inverted, closed-lid measurement approach yields comparable Raman signal of bacteria as the standard upright, open-lid method*

To demonstrate our proposed approach, we collected Raman spectra of bacterial colonies through the bottom of an inverted dish, placed in the same orientation as bacterial culture and storage, and the collected spectra are subject to analysis as outlined in **Fig. 1**. In our approach, referred to as "Inverted" in **Fig. 1A**, the focused light exits from a 10× objective lens and passes through the petri dish and solid agar, comprising approximately 3-4 mm thickness, before reaching target colonies. Upon light-sample interaction, the scattered light travels back the same path. For comparison, we also collected Raman spectra using the standard approach, referred to as "Upright," where light focuses directly onto colonies formed on an agar plate with an open lid. We further assess the signal sensitivity of the proposed approach using *E. coli* colonies with or without genetic modification to express GFP and classify them using machine learning analysis. Notably, as shown in **Fig. 1B**, Raman mappings were generated using the inverted method, which is otherwise not possible in upright, opened plates due to contamination concerns and agar dehydration during the long duration (11h) for high-resolution mapping (1 μm spatial resolution).

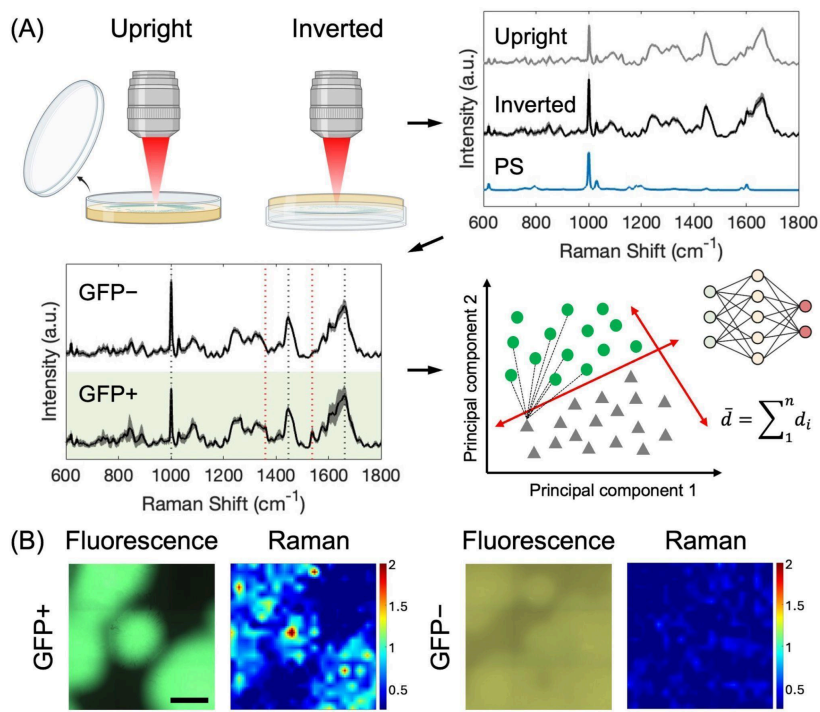

**Figure 1**. Overview of the workflow using the proposed Raman measurement approach for direct bacterial identification using an unopened, solid agar plate placed in an inverted orientation. (A) Typical Raman measurement that allows focused light directly to interact with bacterial colonies exposed to the environment (upright) and our proposed measurement scheme where a culture plate placed inverted with a closed lid, an identical setup as bacterial culture and storage, with light passing through and traveling back through petri dish and agar medium to reach bacterial colonies (inverted) are tested for Raman measurement with *E. coli* colonies. Both upright and inverted approaches show comparable colony signals distinct from the background PS petri dish. Inverted approach shows single-gene level sensitivity that can objectively distinguish *E. coli* colonies based on their GFP expression via quantitative statistical analysis and machine learning analysis. (B) Inverted approach enables time-intensive Raman mapping without perturbing colonies or constant refocusing while preserving signal sensitivity that can distinguish *E. coli* colonies based on their GFP expression, matching with corresponding fluorescent images. Scale bar = 500 μm

We first investigated the inverted approach with *E. coli* colonies prepared on LB agar in a quartz-bottomed petri dish, as quartz is a widely used crystalline substrate due to its weak Raman activity, circumventing interference with the sample specimen[31], and compared with Raman spectra collected via the upright approach. Despite the presence of additional layers to reach the specimen, the inverted approach presented Raman spectra featuring peaks characteristic to *E. coli* colonies at 1002, 1447, and 1662 cm$^{-1}$ that are attributable to phenylalanine, $CH_2$ bending of protein and lipids, and Amide I, respectively (**SFig. 1**)[32–34].

We then investigated commonly used substrate conditions for bacterial culture: PS petri dishes filled with LB agar (**Fig. 1A**) and TSA agar (**SFig. 1-2**). For both conditions, we identified the characteristic peaks of *E. coli* colonies, despite the difference in media background signals. In the

Raman spectra collected via the inverted approach, an increase in peak height at 1002 cm$^{-1}$ was observed for these two conditions, compared to their upright approach counterparts. We believe this is due to PS background signals overlapping with *E. coli* colonies (**SFig. 1**).

*Inverted measurement approach can objectively distinguish bacterial colonies with differences at a single gene level*

To demonstrate the sensitivity of our proposed inverted approach for broad microbial and clinical applications, we compared Raman spectra of colonies of genetically modified *E. coli* tagged with GFPmut3 to the colonies of wild type *E. coli*, hereby referred to as GFP+ and GFP−, respectively. As shown in **Fig. 2** and **SFig. 1**, Raman peaks at 1361 and 1538 cm$^{-1}$ were visibly distinguished in the mean spectra unique to GFP+ colonies using both inverted and standard upright approaches. These peaks agree well with the previous report of GFP base structure[35], confirming the sensitivity of the inverted approach.

The collected Raman spectra were subjected to principal component analysis (PCA) to objectively distinguish GFP+ and GFP− *E. coli* colonies, resulting in a clear separation between the two regardless of the measurement approach (**Fig. 2**). For all three substrate conditions, PCA score distribution plots were analyzed with the 95% confidence interval of GFP+ and GFP− principal component (PC) scores, represented as ellipses showing no overlaps between the two clusters (**Fig. 2A-B** and **SFig. 3**). To objectively assess the clusters, we calculated the Euclidean distance between each PC score between GFP+ and GFP− scores (**SFig. 4**). For both upright and proposed inverted approaches, the intra-group and inter-group distance of PC scores between GFP+ and GFP− colonies presented statistically significant differences, indicating that the two groups can be objectively characterized regardless of the plate orientation as well as background setups.

In the respective PCA loading plots (**Fig. 2C-D**), we identified multiple peaks contributing to the separation, including 1361 and 1538 cm$^{-1}$, matching with the peaks identified via visual inspection of the mean Raman spectra and prior report on GFP structure[35]. Compared to the Raman peaks observed in both GFP+ and GFP− colonies, such as 1002, 1447, and 1662 cm$^{-1}$, the intensity of the GFP-associated peaks in PC loadings was elevated and unambiguously observed in both substrate conditions, demonstrating that PCA accurately captured the contributions of GFP expression in colonies. This observation is corroborated in the respective PCA score distribution plots. In the colonies prepared in LB agar-filled PS dish setup, PC2 appeared as the main axis of the separation between the GFP+ and GFP− clusters (**Fig. 2A**) with PC2 loading spectrum featuring GFP-specific peaks (**Fig. 2C**). Similarly, in the TSA-agar filled PS dish condition, the separation is primarily along the axis of PC3 (**Fig. 2B**) and thus PC3 loading spectrum featuring greater intensity of the GFP-specific peaks than PC2 signal (**Fig. 2D**). Furthermore, we performed a density functional theory (DFT) calculation of the ground-state Raman spectrum for GFPmut3 chromophore, which confirmed our observed peaks characteristic of GFP and identified additional peaks, including 1092 and 1603 cm$^{-1}$. While these additional peaks were not well distinguished in the mean Raman spectra of colonies, potentially due to overlaps with *E. coli* colonies' broad bands, their contributions match with PC loadings. This result indicates that our approach not only provides uncompromised spectral quality but also can be used to objectively distinguish genetic modifications made in bacterial colonies.

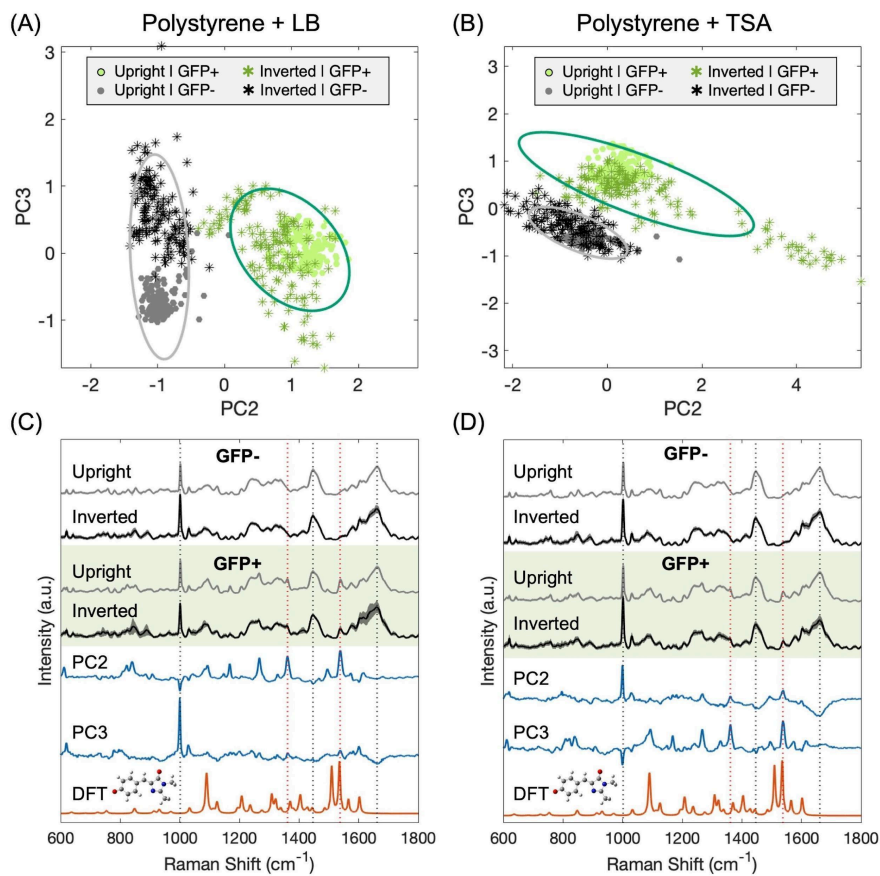

**Figure 2**. PCA results on the Raman spectra of GFP+ and GFP− *E. coli* colonies collected via both upright and inverted approaches. (A-B) PCA score distribution plots and (C-D) corresponding PC loadings and ground-state Raman spectrum for GFPmut3 chromophore estimated by DFT calculation compared with mean spectra of *E. coli* colonies prepared on LB agar-filled (A, C) and TSA agar-filled PS petri dish (B, D). Peaks at 1002, 1447, and 1662 cm$^{-1}$, featured as gray vertical lines, appear for both GFP+ and GFP− colonies, and the peaks at 1361 and 1538 cm$^{-1}$, featured as red vertical lines, are observed only in GFP+ colonies, indicating their association with GFPmut3 chromophore.

Moreover, we performed machine learning analysis using three different models, support vector machine (SVM), random forest (RF), and fully connected network (FCN), for objective classification of GFP+ and GFP− colonies based on their Raman spectra, as shown in **Fig. 3**. First, we investigated classification performance between GFP+ and GFP− colonies for individual setups used in this study, LB agar-filled quartz bottomed and PS petri dish, and TSA agar-filled PS petri dish measured in both orientations (**SFig. 5A**). All three models have shown over 95% classification accuracy for identifying colonies based on their GFP expression. We then tested the classification performance dependence on the spectral data source (measured upright or inverted) used in training versus testing (**Fig. 3A** and **3C**). Classification models performed better when training data was collected via the inverted approach and tested with data collected via upright approach than the reverse, achieving 99.5±0.5% and 88.7±8.7%, respectively, for all three substrate conditions. Furthermore, models trained with the inverted approach showed robust performance varying by less than a mean of 1% across classification models and 0.5% across different substrates, compared to 5% and 17.4%, respectively. This superior classification performance when training with data from the inverted approach may have resulted from the additional noise embedded in the spectra from the media and dish in the light path, enabling the model to learn noise-invariant features, thus making robust classification. On the other hand, the models trained with spectra measured via the upright approach, which focused directly on colonies, cannot make reliable predictions on the spectra influenced by background noise.

We further investigated classification performance independent of the Raman measurement approach by including spectra of all three substrate conditions measured in the same dish orientations for training (**Fig. 3B** and **3D**). Here, compared to predictions from models trained with a single substrate condition, a noticeable increase in overall classification performance was observed for both dish orientations, with 99.2±1.2% and 96.1±4.3% accuracy when trained with inverted and upright measurement data, respectively. Similar to observations made with models trained with a single substrate and measurement scheme, all three classification models trained with spectra measured via inverted approaches achieved higher accuracy than the models trained with upright measurements for colonies prepared in PS dishes - LB-filled PS dishes with an accuracy of 99.9±0.2% and 96.9±3.2% and TSA-filled PS dishes with an accuracy of 100% and 92.0±4.0%, when trained with inverted and upright measurement data, respectively.. Furthermore, the training with upright measured data showed greater substrate dependence in the classification performance, resulting in 97.3-100% for LB-filled quartz-bottomed dishes, which contain the least background contributions, and significantly lower 90.3-99.3% and 85.7-95.7% for LB-filled and TSA-filled PS dishes, respectively, across all three models. However, classification models trained with inverted measurements showed little to no variation in the prediction, resulting in 97.8±1.1%, 99.9±0.2%, and 100±0.0% for LB-filled quartz-bottomed, LB-filled PS, and TSA-filled PS dishes, respectively, relative to the changes in the substrate as well as differences in model algorithms, demonstrating robustness of the performance.

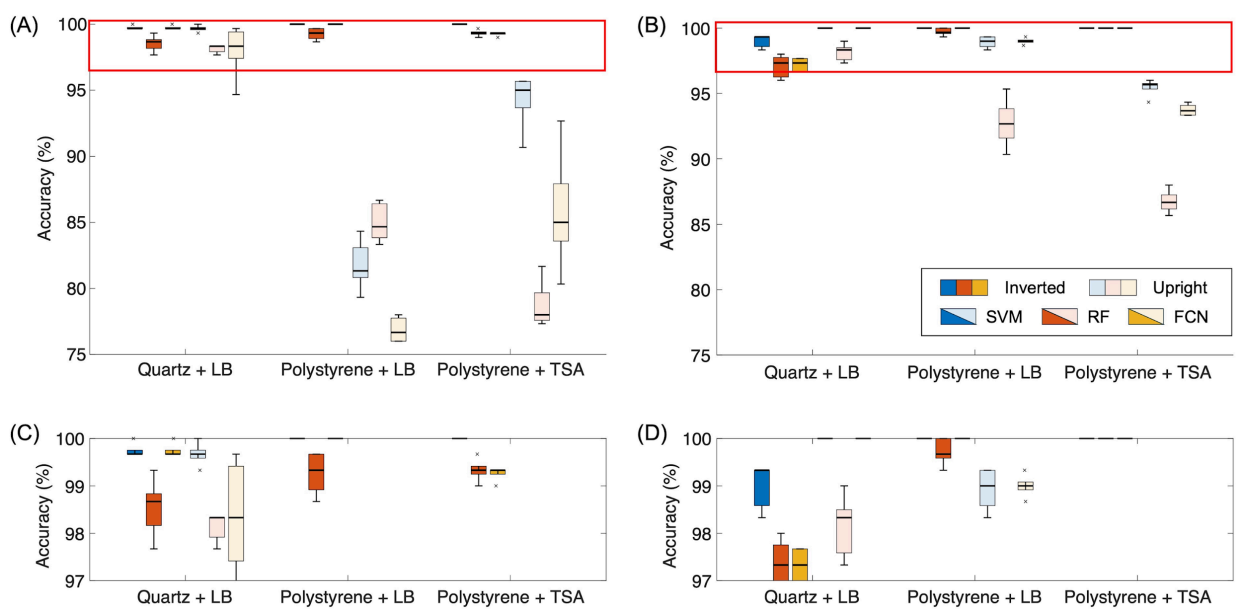

**Figure 3**. Machine learning classification results based on SVM (blue), RF (orange), and FCN (yellow) models to investigate the influence of Raman measurement approaches on identifying between GFP+ and GFP− *E. coli* colonies, subject to different algorithms and background materials. (A) All three models were individually trained with Raman spectra of colonies prepared on LB agar-filled quartz-bottomed petri dishes (Quartz + LB), LB agar-filled PS petri dishes (Polystyrene + LB), or TSA agar-filled PS petri dishes (Polystyrene + TSA) that were measured in either inverted (dark-colored) or upright (light-colored) approaches and tested with Raman spectra of colonies prepared on the same conditions as training but with different measurement approaches. When trained with the inverted approach and tested with the upright approach, mean classification performance across all classification models resulted as 99.3, 99.8, and 99.5% for Quartz + LB, Polystyrene + LB, and Polystyrene + LB, respectively. On the other hand, when trained with the upright approach and tested with the inverted approach, mean classification performance for all classification models dropped for all substrate conditions, resulting in 98.6, 81.2, and 86.3% for Quartz + LB, Polystyrene + LB, and Polystyrene + LB, respectively. (B) All three models were individually trained with Raman spectra of colonies prepared on all three conditions, measured in either inverted (dark-colored) or upright (light-colored) approaches, and tested with Raman spectra of colonies prepared on each preparation condition with different measurement approaches. Including measurements of various substrate conditions for training improves overall classification performance, resulting in 97.8, 99.9, and 100% (training|testing with inverted|upright approaches; dark-colored) and 99.4, 96.9, and 92.0% (training|testing with upright|inverted approaches; light-colored) for Quartz + LB, Polystyrene + LB, and Polystyrene + LB, respectively. (C)-(D) Close-ups of (A)-(B), respectively, ranging 97-100% classification accuracy. All classification models were tested for 5 replicates.

Next, we performed machine learning analysis to investigate classification capability agnostic to differences in substrate material or media, as shown in **SFig. 5B-C**. While the difference in the substrate materials and measurement orientations contributes varying degrees of background noise to the Raman spectra, most models could make accurate classifications, with mean and standard deviation classification performance of 96.5±4.6 and 92.8±9.3% when trained with the inverted and upright approaches, respectively. This shows that our proposed inverted approach preserves features characteristic to bacterial colonies while retaining sensitivity that enables accurate classification based on GFP expression within the same species, despite variance in added background noise as light travels across multiple layers. In particular, classification models trained with colonies prepared in LB agar-filled quartz-bottomed dishes measured via the inverted approach achieved a mean accuracy of 97.5±3.1% when tested with spectra measured with any of the orientation, media, or substrate conditions. However, training data using the upright approach with similar condition showed slightly lower mean classification performance with greater variance, with a mean and standard deviation of 94.4±6.2%, similar to previous observations made with models trained and tested on identical preparation conditions. Interestingly, models trained with data from the inverted approach consistently performed better when tested with data from of when tested with spectra upright approaches, despite the difference in substrate materials; with 99.8±0.4% versus 88.9±3.8% when training with upright collected data. Likewise, the models trained with varying substrate conditions and measured upright presented greater variations in classification performance than those collected with the inverted approach. We believe this is because machine learning models improve generalization capacity when trained with spectra collected via the inverted approach, as they contain greater background contributions than spectra collected via the upright approach.

Combined with our rigorous machine learning-based analysis pipeline, the proposed inverted measurement scheme demonstrated more reliable and robust performance than the standard upright method in classifying *E. coli* colonies based on their GFP expression. This also shows that machine learning models trained in one condition can make accurate classifications on colonies prepared in other conditions, demonstrating the background-agnostic prediction capability.

*Inverted measurement approach integrated with density functional theory informed machine learning enables longitudinal Raman mapping to localize and distinguish bacterial colonies with single gene level differences*

Raman mappings present rich molecular information featuring the spatial distribution of constituent analytes and their relative contributions within the sample specimen. However, Raman mapping typically requires extended measurement time, which is challenging for an open-lid agar plate setup using the upright approach due to contamination risks and uneven culture quality. Here, we demonstrate that the inverted approach enables Raman mapping directly from solid agar culture plates without perturbation or risking media drying, unlike open-lid, upright approach. As shown in **Fig. 4**, we could retrieve *E. coli* colony signals distinct from the background and distinguish colonies based on their GFP expression, as confirmed by their corresponding brightfield and fluorescence images. The colony signal around 1447±5 cm$^{-1}$ corresponds to the deformation vibrational mode of $CH_2$ scissoring in carbohydrate and lipid[36]. The peaks identified from DFT computation of ground-state Raman spectrum for GFPmut3 chromophore were used to distinguish colonies based on their GFP expression. The difference was featured by the Raman peak ratio of 1538 and 1581 cm$^{-1}$, attributable to GFP and peptidoglycan, respectively[37,38]. While the peak at 1581 cm$^{-1}$ was observed in both GFP+ and GFP− colonies, the peak at 1538 cm$^{-1}$ was observed only in GFP+ colonies, as corroborated by DFT calculation, resulting in a greater signal intensity in the Raman map. Moreover, k-means clustering (KMC) on Raman spectral mapping demonstrated objective colony identification. The KMC-masked Raman mapping at 1447±5 cm$^{-1}$ demonstrated a clear separation of bacterial colonies from the background, presenting its potential application for Raman mapping-based automated colony selection. These observations agree with the analysis of the point-scanned spectra, supporting the robustness of the proposed inverted approach as well as presenting its wide application.

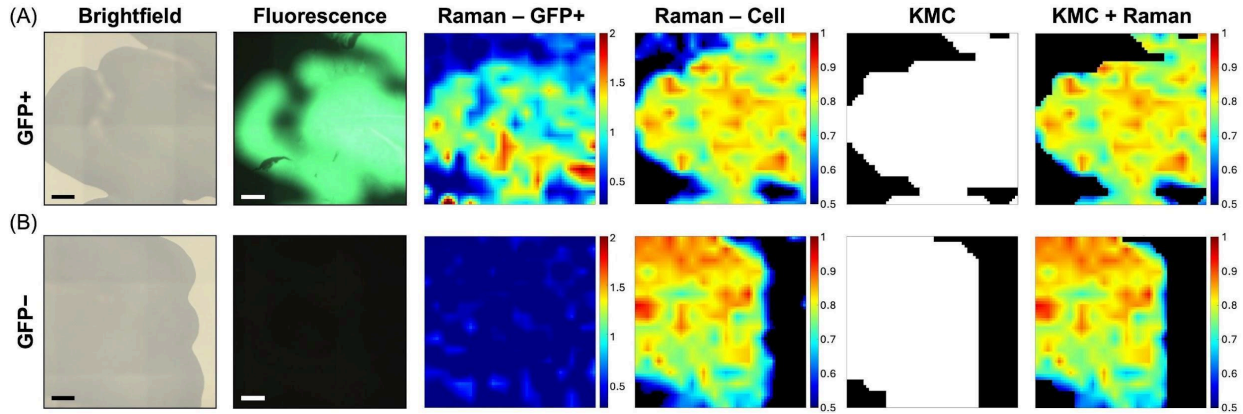

**Figure 4**. Correlated microscopic and Raman mapping images of (A) GFP+ and (B) GFP− *E. coli* colonies obtained via inverted approaches. From left to right: brightfield, fluorescence, Raman mapping at $I_{1538}/I_{1581}$, Raman mapping at $I_{1447\pm5}$, KMC results on Raman mapping, and KMC-masked Raman mapping at $I_{1447\pm5}$ collected from (A) GFP+ and (B) GFP− *E. coli* colonies. Raman maps ($I_{1538}/I_{1581}$) clearly show the GFP expression of colonies, which can be confirmed by matching fluorescence images. KMC distinguishes *E. coli* colony signals from the background, which can be confirmed by matching brightfield images and Raman maps ($I_{1447\pm5}$). Scale bar = 200 μm

**Discussion**

In this work, we demonstrate a direct Raman measurement approach for simple, rapid bacterial identification without perturbing culture conditions or using any extrinsic labels, presenting the potential for seamless integration of Raman spectroscopy in clinical and microbiology laboratory workflows. Unlike labor-intensive standard staining, biochemical testing, and sequencing-based identification that rely on specific enzymes or DNA profiles of bacteria, Raman spectroscopy profiles cells' intrinsic biomolecular properties to characterize bacterial colonies in a non-destructive manner. Our method allows for colony identification without interfering with regular workflows via direct Raman measurement of unopened solid agar culture placed in an inverted dish orientation, promising contamination- and damage-free measurements. We demonstrated the bacterial identification capability of this method with *E. coli* colonies grown on different agar and petri dish materials, showing robust recovery of Raman bacteria signals despite variations in background signals. Furthermore, our approach presented potential for Raman-based identification with single-gene level sensitivity that could distinguish *E. coli* colonies based on their GFP expressions, despite light penetration through 3-4 mm of solid agar and petri dish (**SFig. 6**). We extensively validated our approach through analysis with PCA, machine learning algorithms, and DFT calculations that allowed for objective colony identification.

By maintaining a closed lid and inverted orientation of the sample plate, our method mitigates concerns for contamination or inhomogeneous culture quality. This approach enables the possibility for downstream incubation as well as additional measurement with different assays using the same sample, which is not possible with any other identification methods. Additionally, the closed-lid setup alleviates agar dehydration concerns, which would otherwise require frequent refocusing to optimize Raman signal in between measurements, disturb cell condition, and challenge long-term monitoring of bacterial culture. This approach enabled time-intensive Raman mapping that can objectively identify colonies from the background as well as characterize them based on their GFP expression. While further study with a wide range of bacterial species, strains, and culture materials should be followed, this study holds promise for the wide adoption of Raman spectroscopy in microbiology and clinical workflow for rapid, high-throughput, and objective bacterial identification. It also presents the potential for physical integration of Raman spectroscopy with an incubator for real-time, continuous monitoring of colony formation and antimicrobial resistance acquisition, as well as the immediate to long-term response to antibiotics for improved and timely treatment recommendations.

Moreover, an objective, quantitative assessment of PCA results on Raman spectral data is demonstrated in this study. PCA is one of the most used approaches in analyzing Raman spectra that enables molecular characterization within the sample by decomposing complex spectra into multiple constituents with contributions of each constituent variable[39]. Using PCA, Raman spectra can be subject to classification, as demonstrated in this study for distinguishing GFP+ and GFP− colonies, by projecting PC scores onto respective, orthogonal spaces. Based on the dataset, PC scores form distinct or weak clusters, demonstrating how closely or distantly related each score is to one another in the score domain, and the influence of different variables is represented as PC loadings featuring specific molecular assignments. Our study further advances this analysis approach by quantitatively assessing clusters with respect to how weakly or densely each cluster is formed and their relations to other clusters by calculating a 95% confidence interval and the Euclidean distance between each PC score (**Fig. 2**, **SFig. 3-4**). Through quantitative analysis of PCA results, we objectively interpret Raman spectra of bacterial colonies with differences in GFP expression.

We also demonstrated the capability of machine learning models for classifying Raman spectra of bacterial colonies regardless of their sample preparation conditions. As Raman spectroscopy is sensitive to changes in background artifacts, optimization of substrate conditions and measurement setup is obligatory, particularly for biological specimens being weak Raman scatterers. Thus, machine learning models for analyzing Raman spectra have been typically demonstrated for a uniform substrate condition[34,40–42]. On the other hand, our study demonstrated that classification models that were optimized with one culture condition can make accurate predictions of bacterial colonies grown on different agar or plate materials. This result shows that machine learning models can learn features characteristic to bacterial colonies, despite different background contributions. While we acknowledge only one species is used for this study, the sensitivity of our proposed approach, capable of distinguishing colonies of the same strain based on their GFP expressions, presents its potential for bacterial identification of different strains and species. Our observation of the robust classification capability of various machine learning algorithms invariant to the difference in substrate materials further expands the application of machine learning for analyzing Raman spectra and incorporating publicly available Raman data for generalization.

Our proposed approach will open up a new avenue for direct, label-free bacterial identification and sensitive characterization using Raman spectroscopy in a non-destructive manner while eliminating concerns for potential contamination and extra sample preparation requirements. Additionally, this approach enables longitudinal Raman mapping whilst avoiding issues with agar dehydration, which would otherwise necessitate frequent re-focusing on the colony surface in between measurements as well as lead to inhomogeneous colony quality even within a single specimen. These advantages of our Raman spectroscopy-based direct bacterial colony identification strategy will be valuable across various fields, particularly in clinical diagnostics, microbiome research, and general bacterial research, as it enables real-time, continuous monitoring of colonies without disrupting culture conditions. Along with demonstrating our approach, our work involved a quantitative evaluation of widely used PCA and proposed a robust computational pipeline integrating DFT-based fingerprinting with a machine learning classification strategy for objective analysis of Raman spectra, providing an analysis pipeline for objective spectral characterization of both biological and non-biological materials.

**Methods and Materials**

Cell culture

*E. coli* strain Seattle 1946 cells, purchased from ATCC (#25922 and #25922GFP for GFP− and GFP+ cells, respectively), were used in this study. Cell suspensions were prepared and maintained using standard culture procedures[43]. Briefly, cells prepared in frozen stocks were streaked onto three different plates: 1) quartz (Ø 1 mm)-bottomed petri dishes (Ø 35 mm) filled with solidified LB agar media; 2) PS plates (Ø 100 mm) filled with solidified LB agar media (Teknova, Cat. # L1100); and 3) PS plates (Ø 100 mm) filled with TSA (Hardy Diagnostics, Cat. # G62). Cells were incubated and allowed to grow at 37°C for 24 hours. The Raman spectra of *E. coli* colonies prepared on solid agar plates were collected using two different plate orientations: 1) inverted: a plate placed in an inverted orientation with a closed and sealed lid; 2) upright: a plate placed in an upright orientation with an open lid. The first orientation is identical to the typical placement used for standard bacterial culture and storage. The second orientation is typically used for direct Raman measurement, allowing incident light to directly focus onto the colonies.

Raman Spectroscopy

The Raman spectra of *E. coli* colonies prepared on solid agar plates were collected using a commercial Raman microscope (WITec Alpha300). Briefly, samples were illuminated with a 785 nm laser for 100 s with 1 accumulation at 63 mW (50 mW at the exit of the objective lens). Raman spectra were acquired using a 10× objective (Carl Zeiss; EC Epiplan-Neofluar DIC M27,

NA = 0.90). The collected signals were recorded with a spectrometer (UHTS300 VIS-NIR spectrometer, WITec) connected using multimode optical fiber and thermoelectric cooled CCD (XMC3022-1013) and electron multiplying CCD (EMCCD, Andor DU970N-BV). The CCD and EMCCD are back-illuminated with 1600 × 200 pixels, and each pixel is 16 μm × 16 μm. Collected Raman spectra were subject to pre-processing[31], including cosmic ray removal and denoising, followed by baseline correction[44] and normalization based on the Amide III band.

Data Analysis

The collected Raman spectra were analyzed to identify spectral features characteristic to *E. coli* colonies, distinguished from the background signal from the solid agar and petri dish material. PCA was first performed to decompose Raman spectra into multiple, smaller dimensions to infer the contributions of various molecules from the individual spectrum. PCA was performed with respect to colonies' preparation conditions, including Raman spectra from both upright and inverted measurement schemes of GFP+ and GFP− colonies. Resulting PC scores and respective coefficients were examined for their distribution and characteristic peaks to yield objective identification of bacterial colony signals embedded in Raman spectra. To quantitatively assess clusters formed in the PC score distribution plot, we calculated PC scores of Raman spectra of *E. coli* of the same genetic configurations in both measurement schemes with a 95% confidence interval, represented as ellipses in the PC score distribution (**Fig. 2A-B** and **SFig. 3**). We also evaluated the Euclidean distance of PC scores with respect to GFP presence within the colonies for each Raman measurement scheme (**SFig. 4**).

Raman spectra were further analyzed with three different machine learning models − SVM, RF, and FCN − to test their potential for classifying bacterial colonies with respect to genetic modification, agnostic to the differences in their preparation setup or Raman measurement orientations. Briefly, SVM iteratively searches to optimize a hyperplane that separates Raman spectra into two classes: *E. coli* with and without GFP expression. A linear function was used for optimization. RF is based on an ensemble of decision trees with a maximum number of splits of n-1 and 30 learners, where each tree learns features of the dataset. FCN is the simplest deep learning network consisting of three layers of neurons with ReLu activation in between layers and a softmax activation function to yield output with L1 regularization of 0.0001. The weights of each neuron in the network were initialized by Xavier initialization and optimized up to 1000 epochs. The machine learning models were examined for their classification performance with respect to GFP presence, agnostic to dish orientations during Raman measurement (**Fig. 3**), the preparation conditions, or both (**SFig. 5**). For all three models, we first trained individually with Raman spectra of *E. coli* colonies grown on 1) LB agar-filled quartz-bottomed dishes, 2) LB agar-filled PS dishes, and 3) TSA agar-filled PS dishes (**Fig. 3**). The trained models were then tested with colonies prepared in the same growth condition but collected via the opposite approach, thereby the classification models make predictions on the GFP expression of colonies irrespective of Raman measurement approaches. For example, models trained with inverted approach measurements were tested with upright approach measurements of an identical culture condition and vice versa. All six different combinations of culture and measurement setups were individually trained with the three models − SVM, RF, and FCN − and tested for classification between GFP+ and GFP− *E. coli* colonies. Next, we tested these models to make predictions on colonies prepared in different growth conditions measured via both upright and inverted approaches (**SFig. 5**). In this analysis, we split training and testing sets such that the model optimized for *E. coli* colonies prepared in one condition can make classification of the colonies prepared in the other setups with different background materials. Moreover, we train models with Raman spectra of all growth conditions that were measured using the same approach. These models were tested with colonies prepared in all three growth conditions but collected via the opposite approach, thereby the classification models that were trained with diverse culture setups make predictions on the GFP expression of colonies irrespective of Raman measurement approaches. All analyses were performed and visualized using MATLAB.

DFT Calculations

The geometric structure of the GFPmut3 chromophore is formed from the -Gly65-Tyr66-Gly67- tripeptide motif and further optimized using the B3LYP functional by DFT calculations[45]. 6-311+G** was used as the basis set for C, H, O, and N atoms. The ground-state Raman spectrum for GFPmut3 chromophore was obtained by vibrational calculations with no imaginary frequencies, indicating that the optimized structure was located at the local or global minima of the potential energy surface. The calculated Raman spectrum was smoothed with a half-width at half-maximum of 4 cm$^{-1}$ and further applied with a scaling factor of 0.9679 to better compare with the experimental Raman spectrum[46]. All calculations were conducted using the Gaussian 16 software package[47].

**Acknowledgments**

We would like to thank the insightful suggestions and advice from Dr. Jeon Woong Kang and Dr. Eric Alm, and the initial help from Dr. Jongwan Lee and Anne Hsi-An Huang to optimize experiments. J.H.K. was supported by the MIT Postdoctoral Fellowship for Engineering Excellence.

# Supplementary Information

Non-perturbative Bacterial Identification Directly from Solid Agar Plates Using Raman Micro-spectroscopy and Machine Learning

Jeong Hee Kim[1], Jia Dong[1], Marissa Morales[1], Loza Tadesse[1,2,3,*]

[1]Department of Mechanical Engineering, MIT, Cambridge, MA, USA

[2]Ragon Institute of MGH, MIT and Harvard, Cambridge, MA, USA

[3]Jameel Clinic for AI & Healthcare, MIT, Cambridge, MA, USA

* Corresponding Author:

Loza Tadesse, PhD

Assistant Professor of Mechanical Engineering, MIT
Associate Member, Ragon Institute of MGH, MIT, and Harvard
Associate Member, J Clinic for AI and Health

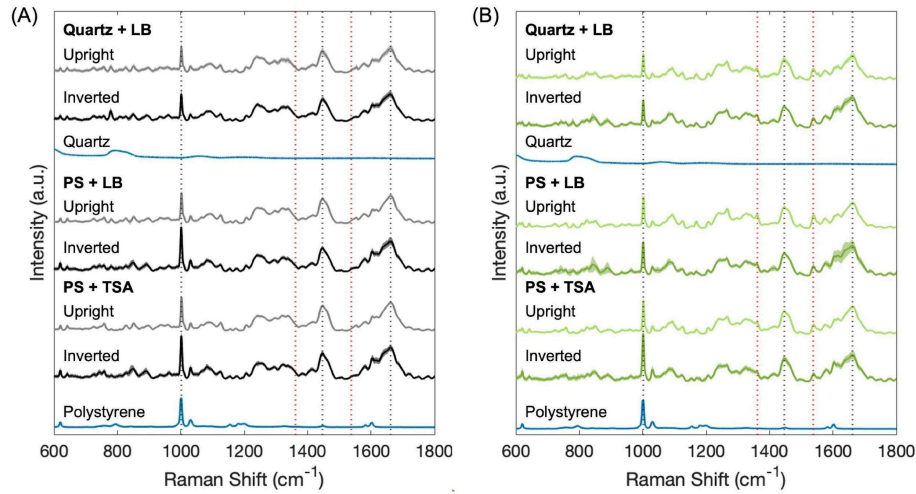

**Supplementary Figure 1**. Mean Raman spectra of *E. coli* colonies collected via upright and inverted approaches (±1 standard deviation shaded). From top to bottom, colonies are prepared on an LB agar-filled quartz bottomed dish (Quartz + LB), an LB agar-filled PS petri dish (PS + LB), and a TSA agar-filled PS petri dish (PS + TSA). Raman spectra feature peaks indicative of *E. coli* colonies at 1002, 1447, and 1662 cm$^{-1}$ (gray dotted vertical lines) for both (A) GFP− and (B) GFP+ colonies, while distinguishing GFP expressions at 1361 and 1538 cm$^{-1}$ (red dotted vertical lines).

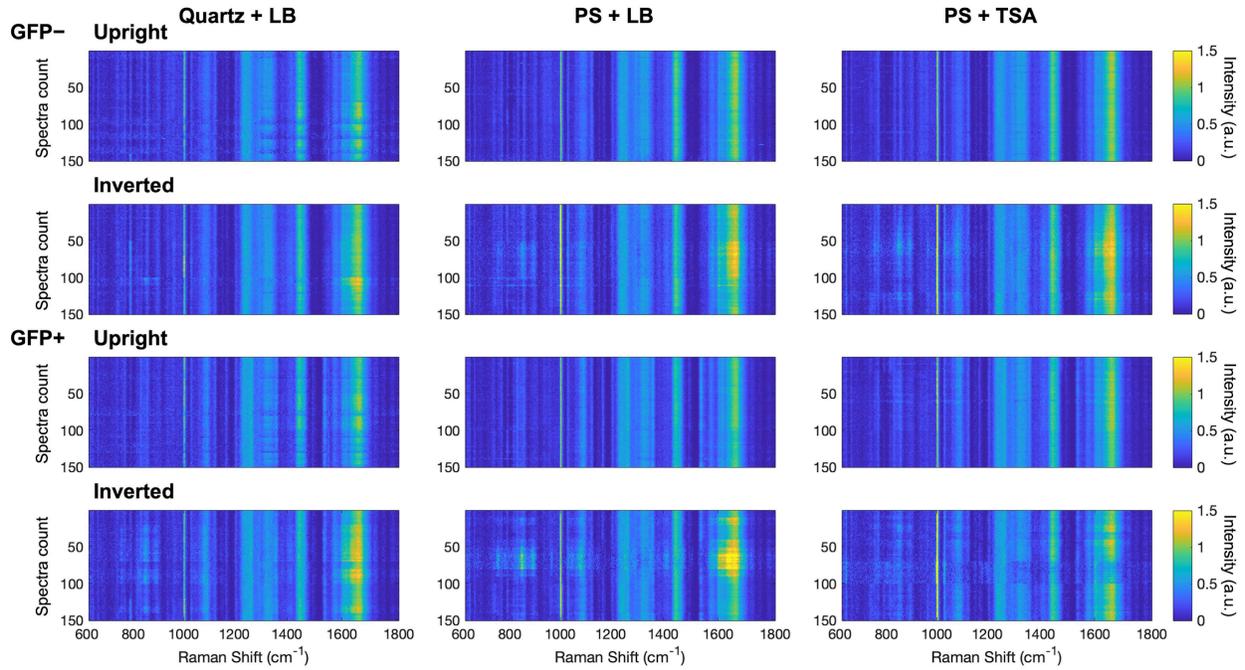

**Supplementary Figure 2**. All Raman spectra of GFP− and GFP+ *E. coli* colonies collected via upright and inverted approaches. From left to right, colonies are prepared on an LB agar-filled quartz-bottomed dish (Quartz + LB), an LB agar-filled PS petri dish (PS + LB), and a TSA agar-filled PS petri dish (PS + TSA).

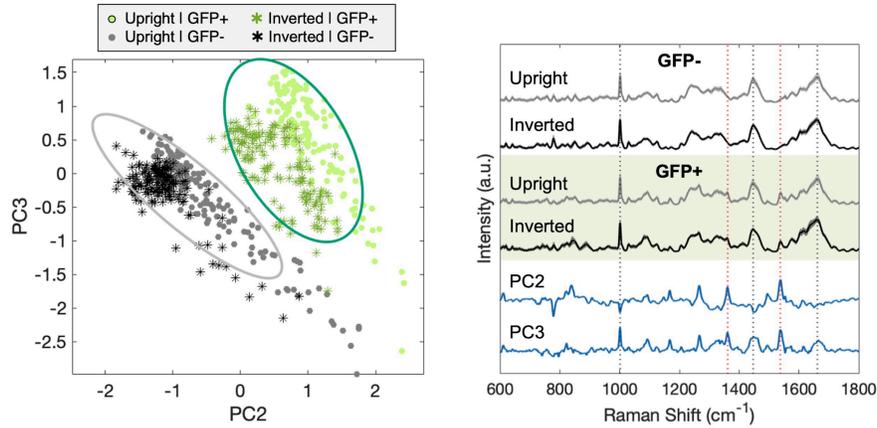

**Supplementary Figure 3**. PCA results on the Raman spectra of GFP− and GFP+ *E. coli* colonies prepared on an LB agar-filled quartz-bottomed petri dish collected via both upright and inverted approaches. PCA score distribution plots (left) and corresponding PC loadings compared with mean spectra of *E. coli* colonies (right).

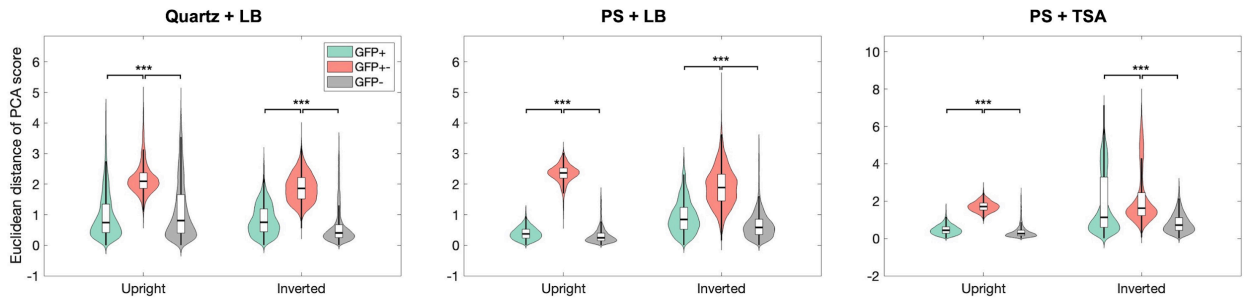

**Supplementary Figure 4**. Euclidean distance between each data point between GFP+ and GFP− PC scores based on PCA of the Raman spectra of *E. coli* colonies collected via both upright and inverted approaches. Each plot indicates the distance of all data points within the group: intra-group distance of all PC scores within GFP+ (green) and GFP− (gray), and inter-group distance between GFP+ and GFP− (red). From left to right, colonies are prepared on an LB agar-filled quartz-bottomed dish (Quartz + LB), an LB agar-filled PS petri dish (PS + LB), and a TSA agar-filled PS petri dish (PS + TSA). P-value: * p≤0.05, ** p≤0.01, *** p≤0.001

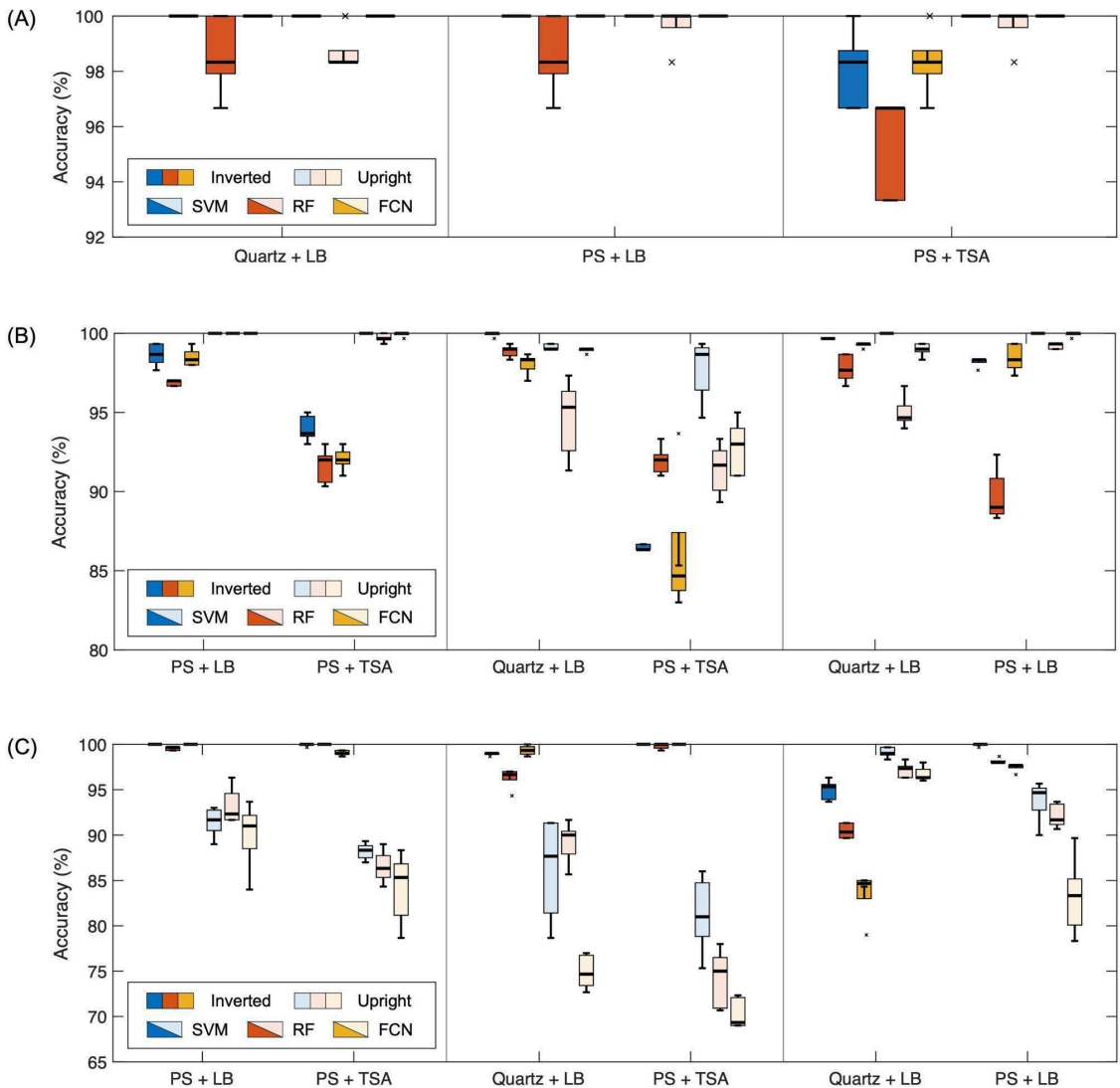

**Supplementary Figure 5**. Machine learning classification results based on SVM, RF, and FCN models to investigate the capacity of identifying between GFP+ and GFP− *E. coli* colonies. All three models were individually trained with Raman spectra of colonies prepared on one of the three conditions: 1) LB agar-filled quartz-bottomed petri dishes, 2) LB agar-filled PS petri dishes, or 3) TSA agar-filled PS petri dishes – measured in either upright or inverted approaches. The trained models were then tested with Raman spectra of the colonies prepared in (A) the same condition as the training, and the other two conditions measured of (B) the same orientation as the training set and (D) different orientation from the training set. (dark-colored: training with inverted and testing with upright measurements; light-colored: training with upright and testing with inverted measurements)

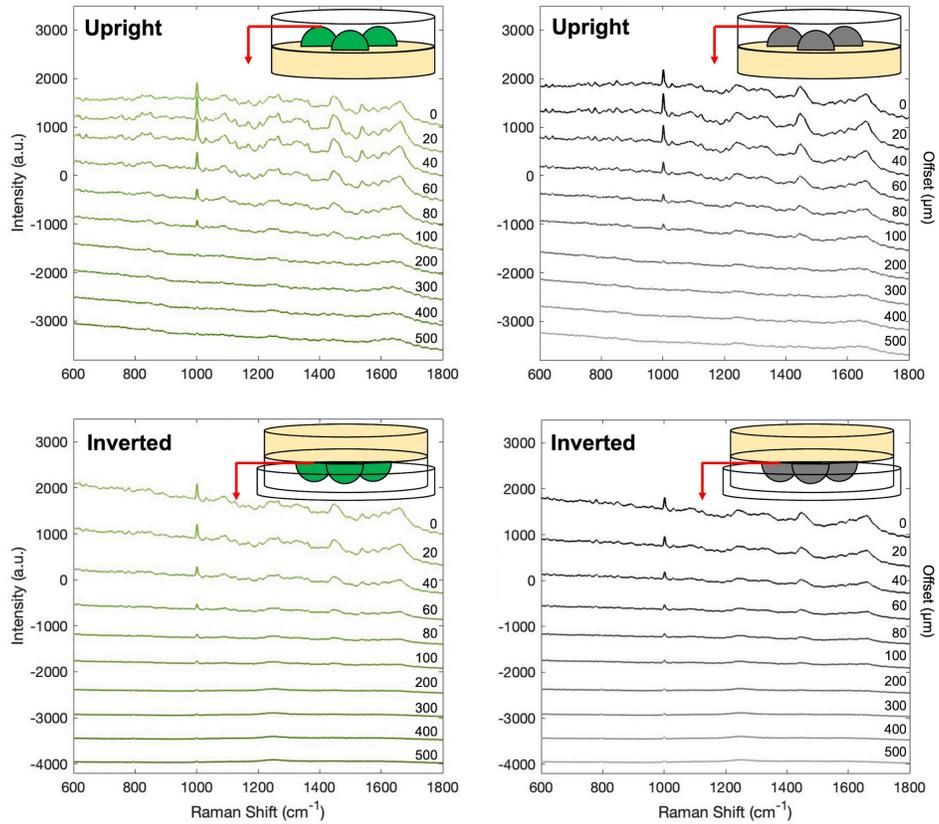

**Supplementary Figure 6**. Mean Raman spectra of GFP+ (left) and GFP− (right) *E. coli* colonies prepared on an LB-agar-filled quartz-bottomed petri dish, collected via both upright and inverted approaches at various depths around the colonies. For the upright approach, features indicative of bacterial colonies sustain up to 40 μm below the colony surface and depreciate as the focal plane moves closer to the agar. For the inverted approach, features indicative of bacterial colonies rapidly drop as the focal plane moves away from the agar. n = 7-9 per spectrum